MD. ABDUL LATIF SARKER, CHONBUK NATIONAL UNIVERSITY, KOREA

# An Error Covariance Splitting Technique for Multi-User MIMO Interference Environment

*Abstract*—This paper investigates an error covariance matrix splitting technique for multiuser multiple input and multiple output (MIMO) interference downlink channel. Most of the related work has thus far considered the traditional error covariance matrix which has not been well-shaped for maximizing the system capacity. Thus, we split and propose a new iterative error covariance matrix to mitigate the system error and maximize the system capacity in this paper. Numerical results illustrate that our proposed method is strictly better than the traditional method.

## 1. Introduction

Multiuser MIMO communication offers high system capacity of the downlink channel which is achieved by a block diagonalization method with joint design of transmit precoding and receive filter matrices for the MIMO interference channel in [1-2]. The block diagonalization method is to pre-removed intra-cell interference at the perfect channel state information (CSI) of the transmitter and accounts for the presence of inter-cell interference because the total interference plus noise covariance matrix can be estimated at each user. Note that the interference suppression matrix is only determined by the interference plus noise covariance matrix independent of each user's channel. Joint transmitter and receiver designs have been well-studied in the literature [3-5]. The joint transceiver schemes for the single-user point-to-point MIMO channel were proposed in [3] and various joint transceiver designs were developed in MIMO downlink channel in [5-8].

Most existing works which minimize the error covariance design problem or maximize mutual information have been addressed for precoding designs in [4, 9-10]. The precoding design techniques for the broadcast channel including BD and DPC have only considered multi-user interference from different users in the same cell, without taking into account inter-cell interference. When the receiver accounts inter-cell interference signal that comes from adjacent cells has been shown to significantly degrade the overall system performance in [11-13]. In [7] proposed a compensated subspace tracking algorithm for the online computation of the outer precoder to solve an optimization problem formulated on Grassmannian manifold. In [10] proposed a joint and iterative transceiver design method for the MIMO interference channel to optimize the system performance under a given and feasible degree of freedom. In [11] proposed an iterative algorithm based on the gradient projection method for the transmit covariance matrix design problem. The algorithm allows each user to update its own covariance matrix locally, provided that the channel state information and the covariance matrices of other users can be gathered. Based on a local linear approximation, in [14] proposed a distributed algorithm which lets each user update its own covariance matrix by solving a convex optimization problem. This algorithm can be viewed as the MIMO extension version of the sequential distributed pricing algorithm in [15].

By fixing the receiver structure to any of the standard linear receivers such as zero-forcing (ZF) of minimum mean square error (MMSE), authors [9] reduced the transceiver design to a transmit coovariance matrix design problem. Minimizing the error variance directly with respect to the precoder is difficult if the closed form MMSE equalizers are applied. Therefore, to overcome this difficulty authors were proposed an iterative linear MMSE method which computes the inner precoder and match filter matrices iteratively in [10]. Consequently, a duality is generated between the precoder and match filter matrices. Thus, we split and propose a new iterative error covariance matrix to overcome an error covariance design problem.

This paper is organized as follows:

First, we investigate the system model in Section II. We then present an intra-cell interference cancellation using block diagonalization method in Section III. In Section IV we investigate the sum-capacity maximization and Problem formulation. After that, we split and propose a new iterative error covariance matrix in Section V. Finally, numerical results and conclusions are presented in Sections VI and VII, respectively.

## 2. System Model

We consider a multiuser MIMO downlink channel $\mathbf{H}_k \in \mathbb{C}^{KM_{R,k} \times M_T}$ with $K$ user each employing $M_{R,k} = 1$ receive antennas for the $k-th$ receiver and receiving their own data streams which are precoded transmit symbol $\mathrm{x}_k = \mathbf{F}_k \tilde{\mathrm{x}}_k$ at the BS with $M_T$ transmit antennas in Fig.1 whereas $\mathbf{F}_k$ is the precoder matrix, $\tilde{\mathrm{x}}_k$ is the transmitted signal of the $k-th$ user with average power $P_k = \mathbb{E}\left\{ \tilde{\mathrm{x}}_k^H \tilde{\mathrm{x}}_k \right\}$, respectively. Using a multicellular interference environment, the received signal of $k-th$ receiver is given by

$$r_k = \underbrace{\mathbf{H}_k \mathbf{F}_k \tilde{\mathrm{x}}_k}_{desired\ signal} + \underbrace{\mathbf{H}_k \sum_{l=1, l \neq k}^{K} \mathbf{F}_l \tilde{\mathrm{x}}_l}_{intra\text{-}cell\ interference\ signal} + \underbrace{n_k}_{\substack{inter\text{-}cell\ interference \\ signal\ plus\ noise}} \tag{1}$$

where $n_k = \mathbf{H}_{I,k} \mathrm{x}_{I,k} + \mathrm{z}_k$ is the inter-cell interference signal plus noise, $\mathrm{x}_{I,k}$ is a $M_{I,k} \times 1$ vector of interfering signals with average power $P_{I,k} = \mathbb{E}\left\{ \mathrm{x}_{I,k}^H \mathrm{x}_{I,k} \right\}$, $\mathbf{H}_{I,k}$ is the $K_{ac} M_{R,k} \times M_{I,k}$ inter-cell interference channel gain matrix and $\mathrm{z}_k$ is the additive Gaussian noise vector i.e., $\mathrm{z}_k \sim \left( 0, \sigma_z^2 \mathbf{I}_{KM_{R,k}} \right)$, respectively.

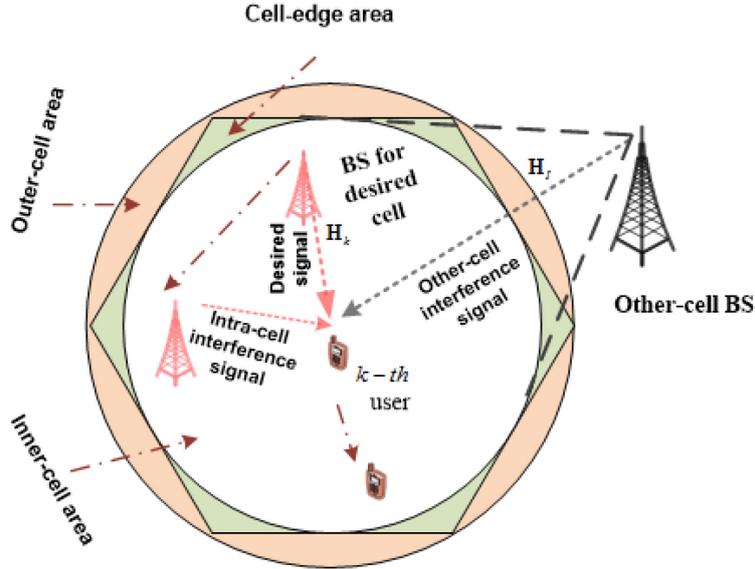

**Fig.1.** Multiuser MIMO system model

## 3. Intra-cell Interference Cancellation Using Block Diagonalization Method

Let, $\mathbf{H}_k \sum_{l=1, l \neq k}^{K} \mathbf{F}_l$ form an effective channel matrix for the $k-th$ user receiver and the $l-th$ user transmit signal $(k, l = 1, ...., K)$. We observe that $\mathbf{H}_k \sum_{l=1, l \neq k}^{K} \mathbf{F}_l$ incurs intra-cell interference to the $k-th$ user unless $\mathbf{H}_k \sum_{l=1, l \neq k}^{K} \mathbf{F}_l = 0_{K_{ac}M_{R,k} \times K_{ac}M_{R,k}}, \forall_{k \neq l}$ in (1) where $0_{K_{ac}M_{R,k} \times K_{ac}M_{R,k}}$ is a zero matrix [16]. Thus, the intra-cell interference free transmission will be warranted as long as the effective channel matrix in (1) can be block diagonalized [xx], that is

$$\mathbf{H}_k \sum_{l=1,l \neq k}^{K} \mathbf{F}_l = 0_{K_{ac}M_{R,k} \times K_{ac}M_{R,k}}, \forall_{k \neq l}, \tag{2}$$

and the equivalent received signal in (1) is now intra-cell interference free, that is,

$$r_k = \mathbf{H}_k \mathbf{F}_k \tilde{\mathbf{x}}_k + n_k \tag{3}$$

## 4. Sum-Capacity Maximization and Problem Formulation

We consider $K_{ac}M_{R,k} \times K_{ac}M_{R,k}$ whitening or an interference suppression filter matrix $\mathbf{W}_k$ to suppress inter-cell interference for each user, then the received signal for $k-th$ user after post-processing is given by (3)

$$\hat{\tilde{\mathbf{x}}}_k = \mathbf{W}_k r_k = \mathbf{W}_k \mathbf{H}_k \mathbf{F}_k \tilde{\mathbf{x}}_k + \mathbf{W}_k n_k = \tilde{\mathbf{H}}_k \tilde{\mathbf{x}}_k + \tilde{n}_k \tag{4}$$

where $\tilde{\mathbf{H}}_k = \mathbf{W}_k \mathbf{H}_k \mathbf{F}_k$ and $\tilde{n}_k = \mathbf{W}_k n_k$.

In the presence of inter-cell interference and using the well-known MMSE receiver in (4), then we can represent the error covariance matrix $\boldsymbol{\varepsilon}_k^{mmse}$ at $k-th$ user receiver as [4, 9]

$$\boldsymbol{\varepsilon}_k^{mmse} = \mathbb{E}\left\{ e_k e_k^H \right\} = \left\{ \mathbf{I}_{K_{ac}M_k} + \tilde{\mathbf{H}}_k \mathbf{Q}_k \tilde{\mathbf{H}}_k^H \boldsymbol{\Upsilon}_{I,k}^{-1} \right\}^{-1} \tag{5}$$

where $e_k = \left( \hat{\tilde{\mathbf{x}}}_k - \tilde{\mathbf{x}}_k \right)$ is the error vector, $K_{ac}$ is active user, $\boldsymbol{\Upsilon}_{I,k} = \mathbb{E}\left\{ \tilde{z}_k^H \tilde{z}_k \right\} = \mathbf{W}_k \left( \mathbf{H}_{I,k} \mathbf{Q}_{I,k} \mathbf{H}_{I,k}^H + \sigma_z^2 \mathbf{I}_{KM_k} \right) \mathbf{W}_k^H$ is the $K_{ac}M_k \times K_{ac}M_k$ covariance matrix of $\tilde{z}_k$, $\mathbf{Q}_k = \mathbb{E}\left\{ \tilde{\mathbf{x}}_k^H \tilde{\mathbf{x}}_k \right\}$ input signal covariance matrix and $\mathbf{I}_{K_{ac}M_{R,k}}$ is the $K_{ac}M_{R,k} \times K_{ac}M_{R,k}$ identity matrix, respectively.

Thus, the maximum sum-capacity of $k-th$ user follows [1]

$$\mathbf{C}_{\max} = \max_{\{\mathbf{Q}_k : Tr(\mathbf{Q}_k) \leq P_k\}} \mathbf{C}_k \tag{6}$$

where the achievable sum-capacity as [9]:

$$\mathbf{C}_k = \sum_{k=1}^{K} \log_2 \det\left\{ \left( \boldsymbol{\varepsilon}_k^{mmse} \right)^{-1} \right\} \tag{7}$$

However, we observe that in (7), the interference suppression matrix $\mathbf{W}_k$ is only determined by the interference plus noise covariance matrix $\left( \mathbf{H}_{I,k} \mathbf{Q}_{I,k} \mathbf{H}_{I,k}^H + \sigma_z^2 \mathbf{I}_{KM_k} \right)$ independent of each user's channel which decompose a large singular value decomposition in [2] or a complex error covariance in [1]. Thus, we propose a new iterative error covariance matrix to easy decompose the spectral radius perfectly for maximizing the sum-capacity in this paper.

## 5. Proposed New Iterative Error Covariance Matrix Construction

When $M_k = 1$, then the error covariance matrix for $K_{ac}$ active user from (6)

$$\boldsymbol{\varepsilon}^{mmse} = \left\{ \mathbf{I}_{K_{ac}} + \tilde{\mathbf{H}} \mathbf{Q} \tilde{\mathbf{H}}^H \boldsymbol{\Upsilon}_I^{-1} \right\}^{-1} \tag{8}$$

Thus, we can first recall **Theorem 1** as follows:

**Theorem 1:** *For $M_k = 1$ and $K_{ac}$ active users, the $K_{ac} \times K_{ac}$ error covariance matrix $\boldsymbol{\varepsilon}^{mmse}$ splits and possess a sum of the two $K_{ac} \times K_{ac}$ approximately circulant matrices that are given by*

$$\boldsymbol{\varepsilon}^{mmse} = \boldsymbol{\varepsilon}_1 + \boldsymbol{\varepsilon}_2 \tag{9}$$

where $\boldsymbol{\varepsilon}_1 = circ(a_0, a_1, ..., a_{K-1})$ is a $K_{ac} \times K_{ac}$ circulant matrix [17-18] and $\boldsymbol{\varepsilon}_2 = \left[ b_{i-j} \right]_{j,i=1}$ is a $K_{ac} \times K_{ac}$ skew circulant matrix [18] whose element is $b_{-j} = -b_{K_{ac}-j}$ for $j = 1, ..., K_{ac} - 1$.

***Proof of Theorem 1:*** Let, the error covariance matrix $\varepsilon$ is a $K_{ac} \times K_{ac}$ Toeplitz matrix and is given by [17, 19]

$$\varepsilon = \mathbb{E}\left\{ ee^{H} \right\} = \begin{bmatrix} t_0 & t_{-1} & t_{-2} & \cdots & t_{-K_{ac}+1} \\ t_1 & t_0 & t_1 & & \\ t_2 & t_{-1} & t_0 & & \vdots \\ \vdots & & & \ddots & \\ t_{K_{ac}-1} & & \cdots & & t_0 \end{bmatrix}. \tag{10}$$

where the error vector $e = \left( \hat{\tilde{x}} - \tilde{x} \right)$ in [9].

The corresponding elements of the error covariance matrix $\varepsilon$ lead to a system for $\varepsilon_1$ and $\varepsilon_2$ as follows [20]:

$$t_{-j} = a_j + b_j , \quad j = 0, 1, ....., K_{ac} - 1 \tag{11}$$

$$t_{K_{ac}-j} = a_j - b_j , \quad j = 1, 2, ....., K_{ac} - 1 \tag{12}$$

We obtain from (11), $a_0 = t_0 / 2$ and $b_0 = t_0 / 2$ when $j = 0$ and $a_0 = b_0$. Using (12) and (11) for $j = 1, 2, ....., K_{ac} - 1$, we have,

$$a_j = \frac{t_{-j} + t_{K_{ac}-j}}{2} \text{ and } b_j = \frac{t_{-j} - t_{K_{ac}-j}}{2} . \tag{13}$$

Thus, we can calculate $\varepsilon_1$ and $\varepsilon_2$ using (11)-(13) and is given by

$$\varepsilon_1 = \begin{bmatrix} a_0 & a_1 & a_2 & \cdots & a_{K_{ac}-1} \\ a_{K_{ac}-1} & a_0 & a_1 & & \\ a_{K_{ac}-1} & a_{K_{ac}-1} & a_0 & & \vdots \\ \vdots & & & \ddots & \\ a_1 & & \cdots & & a_0 \end{bmatrix}, \tag{14}$$

and

$$\varepsilon_2 = \begin{bmatrix} b_0 & b_1 & b_2 & \cdots & b_{K_{ac}-1} \\ -b_{K_{ac}-1} & b_0 & b_1 & & \\ \vdots & -b_{K_{ac}-1} & b_0 & & \vdots \\ \vdots & & & \ddots & \\ -b_1 & & \cdots & & b_0 \end{bmatrix} \tag{15}$$

Therefore, we observe that in (14) and (15) is obviously circulant and skew circulant matrix.

***Example 1:*** Let, the total number of transmit antennas $M_T = 4$, the total number of users $K = 20$ and total active users $K_{ac} = 4$, each user has one received antenna i.e., $M_k = 1$, respectively. Using interference channel with AWGN variance, SINR range is [-10:5:20], the number of data frame 10, the number of data packet 200, QPSK message bit generator, after channel block diagonalization and applying MMSE equalizer, respectively. Using all this codes in MatLab program (v.2016a), the error covariance matrix $\varepsilon^{mmse}$ is given by

$$\varepsilon^{mmse} = \begin{bmatrix} 1.0000 + 0.0000i & 0.6498 - 0.6493i & 0.6599 + 0.7454i & 0.1732 + 0.9510i \\ 0.6498 + 0.6493i & 1.0000 + 0.0000i & 0.6498 - 0.6493i & -0.5785 + 0.7918i \\ 0.6599 - 0.7454i & -0.0264 - 0.9168i & 1.0000 + 0.0000i & 0.6127 + 0.5069i \\ 0.1732 - 0.9510i & -0.5785 - 0.7918i & 0.8127 - 0.5069i & 1.0000 + 0.0000i \end{bmatrix}$$
$$\tag{16}$$

Applying (11)-(13) in (16) and we get,

$$\boldsymbol{\varepsilon}_1 = \begin{bmatrix} 0.5000 + 0.0000i & 0.4115 - 0.8001i & 0.6599 + 0.0000i & 0.4115 + 0.8001i; \\ 0.4115 + 0.8001i & 0.5000 + 0.0000i & 0.6141 + 0.0712i & -0.2761 + 0.8543i; \\ 0.6599 + 0.0000i & 0.2761 - 0.8543i & 0.5000 + 0.0000i & 0.7127; \\ 0.4115 - 0.8001i & -0.6141 - 0.0712i & 0.1000 - 0.5069i & 0.5000 + 0.0000i \end{bmatrix} \tag{17}$$

and

$$\boldsymbol{\varepsilon}_2 = \begin{bmatrix} 0.5000 + 0.0000i & 0.2383 + 0.1508i & 0.0000 + 0.7454i & -0.2383 + 0.1508i; \\ 0.2383 - 0.1508i & 0.5000 + 0.0000i & 0.6141 + 0.0712i & -0.2761 + 0.8543i; \\ 0.0000 - 0.7454i & 0.2761 - 0.8543i & 0.5000 + 0.0000i & -0.1000 + 0.5069i; \\ -0.2383 - 0.1508i & -0.6141 - 0.0712i & 0.1000 - 0.5069i & 0.5000 + 0.0000i \end{bmatrix} \tag{18}$$

However, we would like to analyze the system total rate depending on the variations of the error covariance matrix $\boldsymbol{\varepsilon}^{mmse}$. To improve (6) perfectly, we can construct the new iterative error covariance matrix $\boldsymbol{\varepsilon}^{mmse}(\alpha)$ is given by using the **Lemma 1** and **Theorem 1** of [18]

$$\boldsymbol{\varepsilon}^{mmse}(\alpha) = (\alpha\mathbf{I} + \boldsymbol{\varepsilon}_2)^{-1}(\alpha\mathbf{I} + \boldsymbol{\varepsilon}_1)(\alpha\mathbf{I} + \boldsymbol{\varepsilon}_1)^{-1}(\alpha\mathbf{I} + \boldsymbol{\varepsilon}_2) \tag{19}$$

where $\alpha$ is the number of iterations, $\rho(\boldsymbol{\varepsilon}(\alpha))$ is the spectral radius of $\boldsymbol{\varepsilon}(\alpha)$, the boundary of spectral radius is $\rho(\boldsymbol{\varepsilon}(\alpha)) \leq \sigma(\alpha) < 1, \ \forall_\alpha > 0,$ and

$$\sigma(\alpha) \equiv \max_{\lambda_j \in \lambda(\boldsymbol{\varepsilon}_1)} \left| \frac{\alpha - \lambda_j}{\alpha + \lambda_j} \right| \cdot \max_{\mu_j \in \mu(\boldsymbol{\varepsilon}_2)} \left| \frac{\alpha - \mu_j}{\alpha + \mu_j} \right|. \tag{20}$$

where $\lambda_j$ and $\mu_j$ is the eigenvalues of the $\boldsymbol{\varepsilon}_1$ and $\boldsymbol{\varepsilon}_2$, respectively and the optimal parameter $\alpha^*$ is chosen such that error can be minimized completely (optimal) when $\boldsymbol{\varepsilon}(\alpha^*) = \mathbf{I}$.

## 6. Numerical Results:

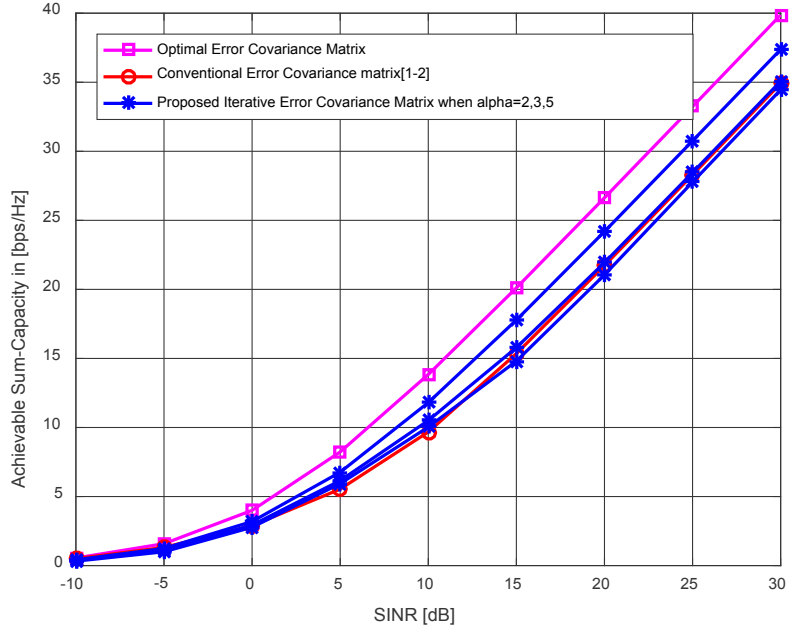

**Fig.1.**

## 7. Conclusions

In this paper, we have investigated a new design of error covariance matrix with interference environment. Many existing works have shown the error covariance matrix design problem with various algorithm in the multiuser MIMO interference environment. Thus, we have split and design a new iterative error covariance matrix which are approximately optimal.

## Acknowledgements

This work was supported by MEST 2015R1A2A1A05000977, NRF, South Korea.